\documentstyle[12pt]{article}
\begin{document}
\begin{center}

{\bf Dynamics of a Quantum Control-Not Gate for an Ensemble of Four-Spin Molecules at Room Temperature}\\ \ \\
{ Gennady P. Berman$^a$, Gary D. Doolen$^a$, Gustavo V. L\'opez$^b$\\and Vladimir I. Tsifrinovich$^c$} 

$^a$Theoretical Division and the CNLS, \\
Los Alamos National Laboratory, Los Alamos, New Mexico 87545\\
$^b$ Departamento de F\'isica, Universidad de Guadalajara,\\
 Corregidora 500, S.R. 44420, Guadalajara, Jalisco, M\'exico\\
$^c$Department of Applied Mathematics and Physics, Polytechnic University,\\
Six Metrotech Center,
Brooklyn NY 11201
\end{center}
\begin{abstract}
We investigate numerically a single-pulse implementation of a quantum Control-Not (CN) gate for an ensemble of Ising spin systems at room temperature. For an ensemble of four-spin ``molecules'' we simulate the time-evolution of the density matrix, for both digital and superpositional initial conditions. Our numerical calculations confirm the feasibility of implementation of quantum CN gate in this system at finite temperature, using electromagnetic $\pi$-pulse.
\end{abstract}
\newpage
Since 1994, Ising spin systems have been considered to be promising candidates for implementing quantum logic gates and for quantum computation \cite{1}-\cite{6}. It was shown in \cite{4,5} that quantum computation can be performed effectively using an ensemble of Ising spin systems at finite temperature. Recent implementations of two-qubit quantum logic gates in an ensemble of spin systems require application of complicated sequences of electomagnetic pulses to the nuclear spins with very close resonant frequencies \cite{4,5}. 
We consider a simple single-pulse implementation of two-qubit quantum logic gate which can be realized in the systems with significant difference between the resonant frequencies. In order to understand the behavior of these systems, it is very important to perform numerical experiments of the dynamics of the single-pulse quantum logic gates and quantum computation for an ensemble of Ising spin systems at room temperature.
Recent suggestions for quantum computation assume the operations using quantum superpositional (entangled) states. The process of manipulating these states in theoretical studies is different from the process of implementation of these states in real physical systems. There are two main problems.  One problem is that these superpositional states are not eigenstates of the corresponding Hamiltonians. These superpositional states constantly change in time. One cannot create time-independent superpositional states even for an isolated system.
 The second problem is that the dynamics of quantum computation involves both resonant and non-resonant interactions. Non-resonant interactions are usually ignored, but they can inhibit the desired effects. For many physical problems (for reasonably small interactions), the resonant dynamics dominates. In this case, the non-resonant dynamics plays an insignificant role which can be reasonably estimated. The situation is completely different for quantum computations because non-resonant effects can accumulate with time, and create significant errors. 

In this paper, we present a numerical analysis of the quantum Control-Not (CN) gate at room temperature. We show that when the initial spin's frequencies differ significantly and the constant of interaction between spins is small enough, the non-resonant effects give small contribution to the dynamics, allowing the single-pulse quantum CN gate to be implemented in this system. 

Quantum CN gates are of the central importance for quantum computation. Any quantum logic gate can be decomposed into a set of one-qubit rotations and CN gates \cite{8}. The two-qubit gate can be described by the operator,
$$
CN=|00><00|+|01><01|+|10><11|+|11><10|.\eqno(1)
$$
The first number, $i$, in $|ij>$ refers to the control qubit, which does not change its value during the CN operation. The second number, $j$, refers to the target qubit, which changes its value only if the control qubit has value ``1''.
The simplest CN gate can be implemented using only one $\pi$-pulse \cite{9}-\cite{11}. It was shown in \cite{11} that, in fact, a $\pi$-pulse provides a modified CN gate, which differs from the ``pure'' CN gate (1) by a phase shift. Nevertheless, the desired feature of the CN gate remains: the $\pi$-pulse changes the state of the target qubit only if the control qubit has value ``1''.

Consider an ensemble of four-spin molecules at finite temperature. Each molecule consists of four spins $I= 1/2$; each spin 
interacts with the other three spins, through an Ising interaction. The whole ensemble can be described by the density matrix, $\rho$. The Hamiltonian of this system, in the reference frame rotating with frequency $\omega$, circularly polarized in $xy$ plane magnetic field, ${\vec h}=h(\cos\omega t,-\sin\omega t)$, is,
$$
{\cal H}=-\hbar\sum_{a=0}^3\Bigg[(\omega_a-\omega)I^z_a+2\sum_{b>a}^3J_{ab}I^z_aI^z_b+
\Omega I^x_a\Bigg].\eqno(2)
$$
In (2), 
$$
\omega_a=\gamma B_a,\quad \Omega=\gamma h.\eqno(3)
$$
where $\gamma$ is the gyromagnetic ratio of each of the four spins; ${\vec B}_a$ is the permanent magnetic field which points in the positive $z$-direction, and is supposed to be different for different spins; $\Omega$ is the Rabi frequency; $J_{ab}$ is the Ising interaction constant. We shall consider the case in which  $J_{ab}=J$. The energy levels for a single molecule containing four spins, for $h=\omega=0$, are shown schematically in Fig. 1. The energy of the ground state is,
$$
E_0=-{{\hbar}\over{2}}\Bigg(\sum_{a=0}^3\omega_a+6J\Bigg).\eqno(4)
$$
The first four excited states in Fig. 1 correspond to the one-spin excitations; the next six states correspond to the two-spin excitations; the next four states correspond to the three-spin excitations, and the last state, $|1111>$, corresponds  to the total inversion of the spin molecule. For the density matrix formalism, it is convenient to use the decimal notation. In this notation we have,
$$
|0000>=|0_30_20_10_0>\rightarrow |0>, \quad 
|0001>=|0_30_20_11_0>\rightarrow |1>, ...,\eqno(5)
$$
$$  
|1111>=|1_31_21_11_0>\rightarrow |15>,
$$
as shown in Fig. 1. Following the idea suggested in \cite{4}, we assume that (using a particular sequence of electromagnetic pulses) an ensemble of four-spin molecules is initially prepared in the state which can be described by the density matrix,
$$
\rho=E/16+\rho_\Delta,\eqno(6)
$$
$$
\rho_\Delta={{\hbar\sum_{k=0}^3\omega_k}\over{2k_BT}}\Bigg[|0><<0|+{{1}\over{2}}\Bigg(-|4><4|+|5><5|+|6><6|+|7><7|+
$$
$$
|8><8|-|9><9|-|10><10|-|11><11|\Bigg)-|12><12|\Bigg].
$$
In (6), $E$ is the unit matrix, and $\rho_\Delta$ is the deviation matrix. The four states, $|k>$, $k=0,1,2,3$ are the ``active states'', which are supposed to be manipulated without noticeable change in the other twelve states. In binary notation, the ``active states'' are $|00ij>$. They are expected to evolve like the pure quantum states of a  two-spin molecule. In Fig. 1, the ``active states'' are marked by a ``bullet''. 

The simplest CN gate for the system under consideration can be realized by a $\pi$-pulse with frequency $\omega_0+3J$ (see Fig. 1). For our case, we shall ``invert'' the convention. Namely, we shall associate with the ground state, $|0>$, of the spin, the value ``1''of the qubit; and the excited state, $|1>$, with the value ``0''. 

A $\pi$-pulse with the frequency $(\omega_0+3J)$ is expected to drive the right spin only if the neighboring spin is in the ground state, i.e.,
$$
|00ij>\rightarrow \cases{|00ij>, &if $i=1$,\cr
|$00$i \bar j>$$, &if  $i=0$,  ($\bar j$=1-$j$),\cr}\eqno(7)
$$
ignoring the overall phase factor. The frequency $(\omega_0+3J)$ is  unique frequency in this system. So, no other states are supposed to be changed by this pulse.

To analyze the dynamics of the CN gate, we first calculate the 16 diagonal and 64 non-zero off-diagonal matrix elements of the Hamiltonian (2). Next, we solve numerically the system of 256 equations of motion for the matrix elements of the density matrix $\rho_\Delta(t)$, for the initial conditions,
$$
\rho_\Delta(0)=
{{\hbar\sum_{k=0}^3\omega_k}\over{2k_BT}}\Bigg[\sum^3_{n,k=0}r_{nk}(0)|n><k|+
{{1}\over{2}}\Bigg(-|4><4|+|5><5|+|6><6|+\eqno(8)
$$
$$
|7><7|+|8><8|-|9><9|-|10><10|-|11><11|\Bigg)-|12><12|\Bigg].
$$
The coefficients $r_{n,k}(0)$ ($0\le n,k\le 3$) in (8) describe arbitrary initial conditions for the ``active part'' of the deviation density matrix, $\rho_\Delta$. Certainly, 
$$
\sum_{n=0}^3r_{nn}=1.\eqno(9)
$$
The deviation density matrix,
$$
\rho_\Delta(t)=
{{\hbar\sum_{k=0}^3\omega_k}\over{2k_BT}}\sum^{15}_{n,k=0}r_{nk}(t)|n><k|,\eqno(10)
$$
with the initial condition (8), was calculated numerically without any further approximations.

The results of these numerical calculations are shown in Fig. 2 and  3, for the following values of parameters,
$$
\omega_k=100(k+1),\quad J=10,\quad \Omega=0.1,\quad \omega=130,\eqno(11)
$$
where $k=0,1,2,3$. The characteristic dimensional parameters can be obtained, for example, by multiplying the parameters in (11) by $2\pi$MHz.
Fig. 2a corresponds to the initial conditions,
$$
r_{00}(0)=1,\quad r_{nk}(0)=0,\quad n,k\le 3,\quad (n,k)\not=(0,0).\eqno(12)
$$
One can see from Fig. 2a, that under the action of a $\pi$-pulse, the coefficient $r_{00}(t)$ (curve 1) decreases monotonically from 1 to 0, indicating the transition,
$$
|0000>\rightarrow |0001>.\eqno(13)
$$
Curves 2 and 3 in Fig. 2a show the evolution of the imaginary and the real parts of the average spin, $Im<I^+>$ and  $Re<I^+>$, correspondingly, where
$$
I^+=\sum_{a=0}^3I^+_a,\quad I^+_a=I^x_a+iI^y_a,\eqno(14)
$$
$$
<I^+>=Tr\{I^+\rho_\Delta(t)\}\sim \sum_{n,k=0}^{15} I^+_{nk}r_{kn}(t).
$$
The value of $Im<I^+>$ describes the precessing amplitude of the spin which is maximum when the average spin is in the $xy$ plane.

Fig. 2b corresponds to the initial conditions,
$$
r_{11}(0)=1,\quad r_{nk}(0)=0,\quad n,k\le 3,\quad (n,k)\not=(1,1),\eqno(15)
$$
and describes the transition,
$$
|0001>\rightarrow |0000>.\eqno(16)
$$
Fig. 2c corresponds to the non-resonant initial conditions,  
$$
r_{22}(0)=1,\quad r_{nk}(0)=0,\quad n,k\le 3,\quad (n,k)\not=(2,2).\eqno(17)
$$
Fig. 2d corresponds to the initial conditions,
$$
r_{33}(0)=1,\quad r_{nk}(0)=0,\quad n,k\le 3,\quad (n,k)\not=(3,3).\eqno(18)
$$
For both cases, shown in Fig. 2c and Fig. 2d, the population of the ``active states'', $|00ij>$, does not change with accuracy of $10^{-3}$. The same is true for the population of all other ``passive states''. 

Now we assume that under the action of a sequence of electromagnetic pulses
we have a superpositional quantum state of a two-spin molecule,
$$
\Psi(0)=c_{00}(0)|00>+c_{01}(0)|01>+c_{10}(0)|10>+c_{11}(0)|11>=\eqno(19)
$$
$$
c_{0}(0)|0>+c_{1}(0)|1>+c_{2}(0)|2>+c_{3}(0)|3>.
$$
For an ensemble of four-spin molecules at room temperature the
superpositional state (19) corresponds to the initial values of $r_{nk}(0)$ in (8),
$$
r_{nk}(0)=c^*_n(0)c_k(0),\quad n,k\le 3.\eqno(20)
$$
We investigated the possibility of implementating a
``$\pi$''- pulse quantum CN gate for the ensemble of four-spin molecules.

In Fig. 3 we show the evolution of the diagonal elements, $r_{kk}(t)$, under the action of a $\pi$-pulse, for the initial conditions,
$$
c_0(0)=0.3^{1/2},\quad c_1(0)=0.2^{1/2},\quad c_2(0)=3^{-1/2},\quad c_{3}(0)=6^{-1/2}.\eqno(21)
$$
One can see from Fig. 3, that at the end of a $\pi$-pulse we have: $r_{00}=r_{11}(0)$, $r_{11}=r_{00}(0)$, and the values of $r_{22}$ and $r_{33}$ do not show a noticeable change. The same is true for the ``non-active'' diagonal elements, $r_{44}$,..., $r_{15,15}$. Thus, a ``$\pi$''- pulse provides the implementation of the quantum CN gate for the
superpositional initial conditions (20), (21) as well as for the digital
initial conditions (12), (15), (17), (18).

We conclude that the dynamics of quantum CN gate can be separated into two parts -- resonant and non-resonant, and the non-resonant interaction plays a minor role when the resonant condition $\omega=\omega_0$ is satisfied, and the constant of interaction, $J$, is small enough. In this case, our numerical calculations confirm the  possibility of implementation of 
the quantum CN gate in an ensemble of quantum four-spin molecules, at room temperature, using a $\pi$-pulse. Further numerical investigations are required to analyze the dynamical behavior of the system  as a function of the frequencies, $\omega_k$, in (11). When the frequencies $\omega_k$ in (11) are close to each other, the resonant and non-resonant dynamics are expected to become more mixed. Also, additional numerical experiments are required to analyze the  creation of the initial state (6), and to study the influence of sequence of $\pi$-pulses on the ensemble of spin molecules. These calculations are now in progress.
\section*{ACKNOWLEDGEMENTS}
G.V.L. and V.I.T. are grateful to the Theoretical Division and the CNLS of the Los Alamos National Laboratory for their hospitality.
This work  was partly supported by the Defense Advanced Research Projects
Agency.
\newpage
\begin{center}
{\bf Figure Captions}
\end{center}
\quad\\
Fig. 1. The energy levels for a four-spin molecule. ``Bullets'' indicate the ``active states'' of the effective ``pure'' quantum two-spin system.\\ \ \\
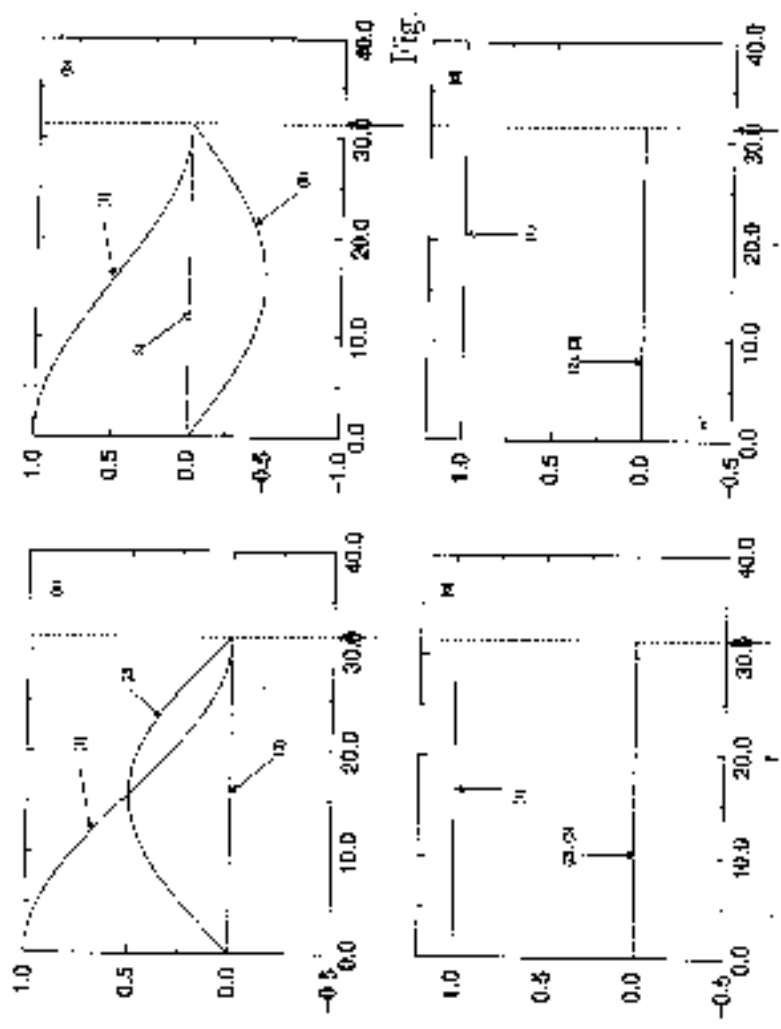
Fig. 2. Evolution of the diagonal density matrix elements, $r_{nn}(t)$, in (10), and the average value, $<I^+>$, for the parameters given in (11).
(a) corresponds to the initial conditions (12); curve (1) refers to $r_{00}(t)$; (b) corresponds to the initial conditions (15); curve (1) refers to $r_{11}(t)$;
(c) corresponds to the initial conditions (17); curve (1) refers to $r_{22}(t)$;
(d)  corresponds to the initial conditions (18); curve (1) refers to $r_{33}(t)$. In Figs 2a-2d curve (2) refers to $Im(<I^+>)$, and 
the curve (3) refers to $Re(<I^+>)$. Vertical arrows indicate the end of the $\pi$-pulse.\\ \ \\
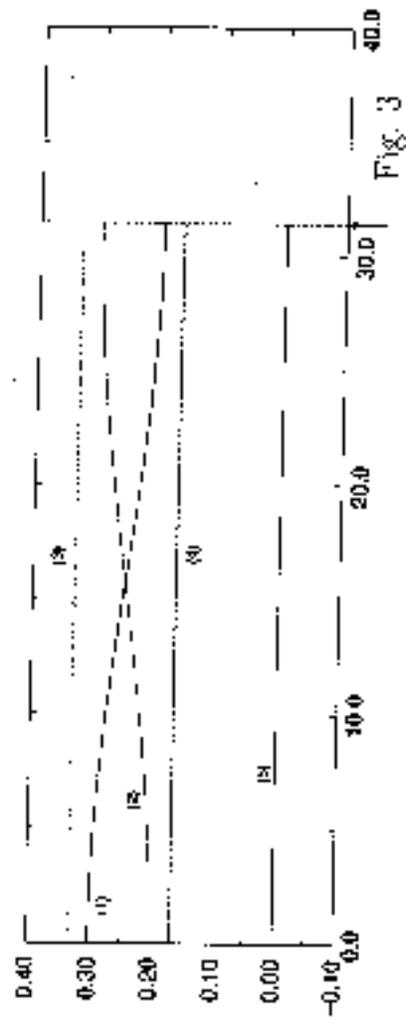
Fig. 3. Evolution of the diagonal density matrix elements, $r_{nn}(t)$, in (10) for the parameters given in (11), and for the superpositional initial conditions (20)-(21). Curve (1) refers to $r_{00}(t)$; curve (2) refers to $r_{11}(t)$; curve (3) refers to $r_{22}(t)$; curve (4) refers to $r_{33}(t)$. Vertical arrow indicats the end of the $\pi$-pulse.
\newpage
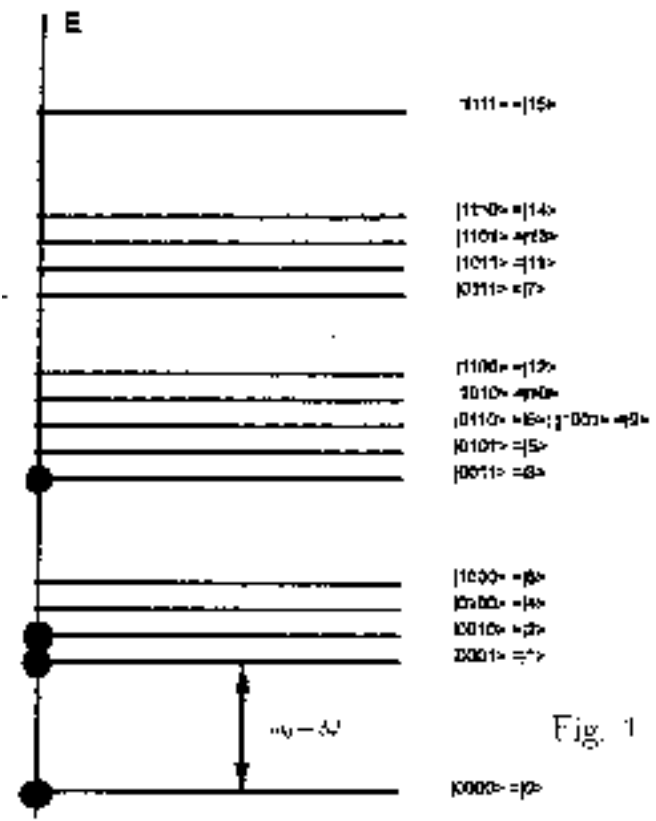
\end{document}